\documentclass[aps,prl,twocolumn,showpacs]{revtex4}
\usepackage{graphicx}
\usepackage{amssymb}
\usepackage{amsmath}

\newcommand{ \be}{\begin{equation}}
\newcommand{ \ee}{\end{equation}}
\newcommand{\beq}{\begin{eqnarray}}
\newcommand{\eeq}{\end{eqnarray}}
\newcommand{\bem}{\begin{pmatrix}}
\newcommand{\eem}{\end{pmatrix}}
\newcommand{\bmx}{\begin{array}}
\newcommand{\emx}{\end{array}}

\begin{document}

\title{Fluid-solid phase transitions in 3D complex plasmas under microgravity conditions}

\author{S. A. Khrapak,$^{1,2}$ B. A. Klumov,$^{1,2}$ P. Huber,$^1$  V. I. Molotkov,$^2$ A. M. Lipaev,$^2$ V. N. Naumkin,$^2$ A. V. Ivlev,$^1$ H. M. Thomas,$^1$  M. Schwabe,$^1$ G. E. Morfill,$^1$ O. F. Petrov,$^2$ V. E. Fortov,$^2$ Yu. Malentschenko,$^3$ and S. Volkov$^3$}

\affiliation{$^1$Max-Planck-Institut f\"ur extraterrestrische Physik, D-85741 Garching,
Germany \\$^2$Joint Institute for High Temperatures, 125412 Moscow, Russia \\$^3$Yuri Gagarin Cosmonaut Training Centre, 141160 Star City, Russia}

\date{\today}

\begin{abstract}
Phase behavior of large three-dimensional complex plasma systems under microgravity conditions onboard the International Space Station is investigated. The neutral gas pressure is used as a control parameter to trigger phase changes. Detailed analysis of structural properties and evaluation of three different melting/freezing indicators reveal that complex plasmas can exhibit melting by increasing the gas pressure. Theoretical estimates of complex plasma parameters allow us to identify main factors responsible for the observed behavior. The location of phase states of the investigated systems on a relevant equilibrium phase diagram is estimated. Important differences between the melting process of 3D complex plasmas under microgravity conditions and that of flat 2D complex plasma crystals in ground based experiments are discussed.
\end{abstract}

\maketitle

\section{Introduction}

Complex (dusty) plasmas -- systems consisting of highly charged micron-size particles in a neutralizing plasma background -- exhibit an extremely rich variety of interesting phenomena ~\cite{Book,PielPPCF,FortovUFN,FortovPR,Ishihara,ShuklaRMP,MorfillRMP,ChaudhuriSM}. Amongst these transitions between fluid and solid phases (freezing and melting) are of particular interest~\cite{Book,FortovUFN,FortovPR,ShuklaRMP,MorfillRMP,ChaudhuriSM,Thomas1994,Chu1994,Hayashi1994,Melzer1994,Morfill_Nat,melt_freez}.
This is largely a consequence of the fact that high temporal and spatial resolution allows us to investigate these phase changes along with various related phenomena at the individual particle level~\cite{Book,FortovPR,MorfillRMP,ChaudhuriSM,Morfill_Nat,melt_freez,MorfillCPP,MorfillPS,Milenko,Knapek1,Knapek,Mitic,Couedel,Nosenko}. Fully resolved atomistic (undamped) particle dynamics provides new insight into natural atomic and molecular systems, whose dynamics cannot be resolved in such detail.

In this paper we report experimental investigations of the fluid-solid phase transitions in large 3D complex plasmas performed under microgravity conditions onboard the International Space Station (ISS). These phase changes are driven by manipulating the neutral gas pressure. Detailed analysis of complex plasma structural properties allows us to quantify the extent of ordering and accurately determine the phase state of the system. Evaluation of various freezing and melting indicators gives further confidence regarding the phase state. It is observed that the system of charged particles can exhibit melting upon increasing the gas pressure, in contrast to the situation in ground-based experiments where plasma crystals normally melt upon {\it reducing} the pressure. This illustrates important differences between generic (e.g. similar to conventional substances) and plasma-specific mechanisms of phase transitions in complex plasmas.

First results from the studies described here have been reported in Refs.~\cite{CrystAIP,CrystPRL}. The purpose of this paper is to provide more detailed and comprehensive information on the experimental procedure, analysis of the obtained results, and their theoretical interpretation.

\section{Experimental Setup}

\begin{figure}
\includegraphics[width=8.2cm]{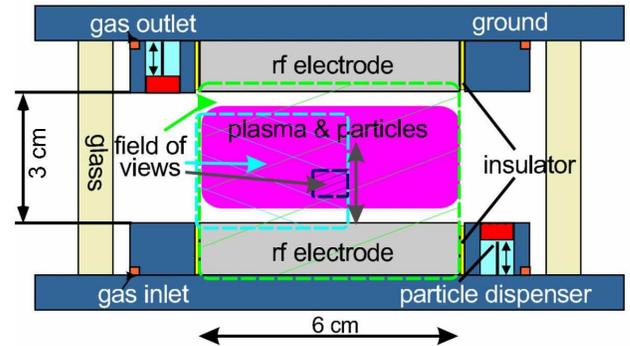}
\caption{(Color online) Sketch of the PK 3 Plus plasma chamber~\cite{PK3+}.}
\label{sketch}
\end{figure}

Experiments are performed in the PK-3 Plus laboratory onboard the ISS~\cite{PK3+}. The heart of this laboratory is a parallel-plate radio-frequency (rf) discharge operating at a frequency of 13.56 MHz, sketched in Fig.~\ref{sketch}. The electrodes are circular plates with a diameter 6 cm made of aluminium. The distance between the electrodes is 3 cm. The electrodes are surrounded by a 1.5 cm wide ground shield including three microparticle dispensers on each side. The dispensers are magnetically driven pistons filled with monodisperse particles of various size and material~\cite{PK3+}.

Discharge can operate in argon, neon or their mixture in a wide range of pressures, rf-amplitudes and rf-powers. The working pressures are in the range between 5 and 255 Pa with and without gas flow. The latter can be produced by a specially designed system, allowing us to operate with the lowest flow rates necessary to provide clean experimental conditions. The flow is essentially a symmetrical gas curtain around the electrode system flowing from the lower to the upper part (see Fig.~\ref{sketch}).

Complex plasmas are formed by injecting monodisperse micron-size particles into the discharge. The optical particle detection system consists of a laser illumination system and three video cameras (there is also a fourth camera of the same type which is used to observe glow characteristics of the plasma). Two diode lasers ($\lambda=686$ nm) are collimated by a system of several lenses producing a laser sheet perpendicular to the electrodes with different opening angles and focal points. Three progressive CCD-cameras detect the reflected light at $90^{\circ}$. An overview camera has a field of view (FoV) of about $60 \times 43$ mm$^2$ and observes the entire field between the electrodes. A second camera has a FoV of about $36\times 26$ mm$^2$ and observes the left part of the interelectrode space (about half of the entire system). The third camera is the high resolution camera with a FoV of about $8\times 6$ mm$^2$. It can be moved along the central axis in the vertical direction. The cameras and lasers are mounted on a horizontal translation stage allowing a depth scan through, and, therefore, 3D observation of complex plasma clouds.

Further details on the PK-3 Plus project can be found in a comprehensive review~\cite{PK3+}.

\section{Experimental procedure}

\begin{figure}
\includegraphics[width=8.2cm]{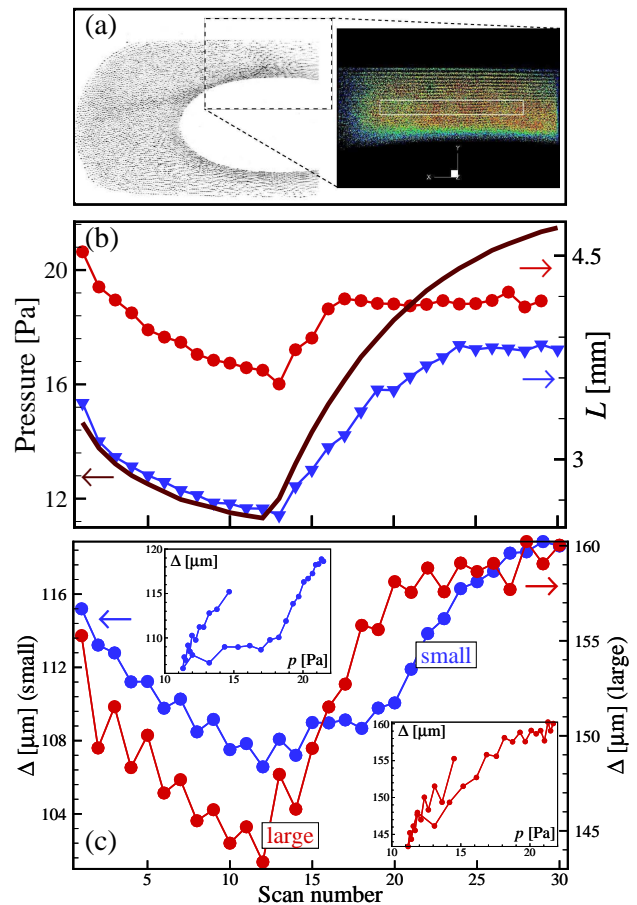}
\caption{(Color) (a) Side view of the particle cloud (inverted colors) taken with the overview camera (left) and the corresponding FoV of the high resolution camera (right) [particles are color-coded to see solid-like (red) and liquid-like (blue) domains]. Rectangle marks the part of the cloud used for the detailed structural analysis (rectangular box $7.0\times 0.7 \times 4.5$ mm$^3$).
(b) Thickness of the particle cloud in the vertical direction vs. the scan number. Blue triangles (red circles) connected by lines correspond to the system of small (large) particles. The corresponding values of pressure are shown by a brown solid curve (the dependence of pressure on the scan number is almost identical in the two runs). The plateaus on the width-curves seen when the pressure increases (right side of the figure) correspond to the situation when the upper cloud boundary leaves the FoV of the high resolution camera; the actual width of the cloud is somewhat larger than shown. (c) Mean interparticle separation $\Delta$ (in the part of the cloud chosen for the analysis) vs. the scan number in the two experimental runs. Blue (red) color corresponds to the system of small (large) particles.  Insets show the dependence $\Delta(p)$ demonstrating some hysteresis, which is more pronounced for small particles.}
\label{boundaries}
\end{figure}

The experiments described here are carried out in argon at a low rf-power ($\sim 0.5$ W). We use two different sorts of particles in the two distinct experimental runs: SiO$_2$ spheres with a diameter $2a =1.55$ $\mu$m and Melamine-Formaldehyde spheres with a diameter $2a =2.55$ $\mu$µm. The experimental procedure, identical in these two runs, is as follows: When the particles form a stable cloud in the bulk plasma, the solenoid valve to the vacuum pump is opened, which results in a slow decrease of the gas pressure $p$. Then, the valve is closed and the pressure slowly increases due to the gas streaming in. (Neutral flow has negligible direct effect on the particles). During the pressure manipulation ($\simeq 6$ minutes in total), the structure of the particle cloud is observed. The observations cover the pressure range from $p\simeq 15$ Pa, down to the lowest pressure of $p\simeq 11$ Pa and then up to $p\simeq 21$ Pa [see Fig.~\ref{boundaries}(a)].

In order to get three-dimensional particle coordinates, 30 scans are performed. Scanning is implemented by simultaneously moving laser and cameras in the direction perpendicular to the field of view with the velocity 0.6 mm/s. Each scan takes $\simeq 8$ s, resulting in the scanning depth of $\simeq 4.8$ mm; the interval between consecutive scans is $\simeq 4$ s. The particle positions are then identified by tomographic reconstruction of the 3D-pictures taken with the high resolution camera observing a region $8\times 6$ mm$^2$ slightly above the discharge center.

\section{Analysis}

\subsection{Global reaction}

Let us first analyze the global reaction of the particle cloud on the pressure manipulation. An example of the particle cloud as seen by the overview and high resolution cameras is shown in Fig.~\ref{boundaries}(a). Figure~\ref{boundaries}(b) shows the cloud thickness in the vertical direction as a function of the scan number (time) for both the systems of small and large particles. It is observed that the position of the upper boundary is strongly correlated with pressure: It moves downwards (upwards) with the decrease (increase) of $p$. This has a clear physical explanation. Particles cannot penetrate in the region of strong electric field (sheath) established near the upper electrode. The position of the upper cloud boundary is thus set by the sheath edge. The sheath thickness is roughly proportional to the electron Debye radius $\lambda_{{\rm D}e}$ which exhibits the following approximate scaling $\lambda_{{\rm D}e}\propto n_e^{-1/2}\propto p^{-1/2}$, where $n_e$ is the electron density. This implies that upon a decrease in the pressure, the particles are pushed farther to the electrode and vice versa, in full agreement with the observations.

The lower cloud boundary, associated with the presence of the particle-free region (void) in the central area of the discharge \cite{Morfill99,GoreeVoid,Comment,Kretschmer,LipaevPRL}, shows less systematic behavior. Its position does not change when the pressure decreases, but then moves slightly upwards when it increases. However, the displacement amplitude is relatively small. As a result, the thickness of the cloud exhibits pronounced decrease (increase) when the pressure decreases (increases), as can be seen in Fig.~\ref{boundaries}(b). The plateaus on the cloud width-curves seen when the pressure increases are artifacts due to the upper cloud boundaries leaving the FoV of the high resolution camera (so that the actual width of the clouds is somewhat larger than shown). Thus, the particle component becomes compressed by reducing the pressure and expands when the pressure is increased. The resulting dependence of the mean interparticle distance (in the part of the particle cloud subject to detailed analysis) on the scan number/pressure is shown in Fig.~\ref{boundaries}(c). The mean interparticle distance $\Delta$ is clearly correlated with pressure, although some hysteresis (more pronounced for the system of small particles) is evident from insets in Fig.~\ref{boundaries}(c).

\subsection{Structural properties}

The observed clouds of particles are not very homogeneous. For example, typical interparticle separations in peripheral regions close to the cloud boundaries can exceed those in the central part of the cloud by a factor of about two. For this reason, a relatively small central part of the cloud sketched in Fig.~\ref{boundaries}(a) has been chosen for the detailed analysis of the structural properties. This part is sufficiently small (especially its vertical extent) so that the system inside is reasonably homogeneous. At the same time, it contains enough particles ($\simeq 10^4$) to yield reasonable statistics.

\begin{figure*}
\includegraphics[width=17.7cm]{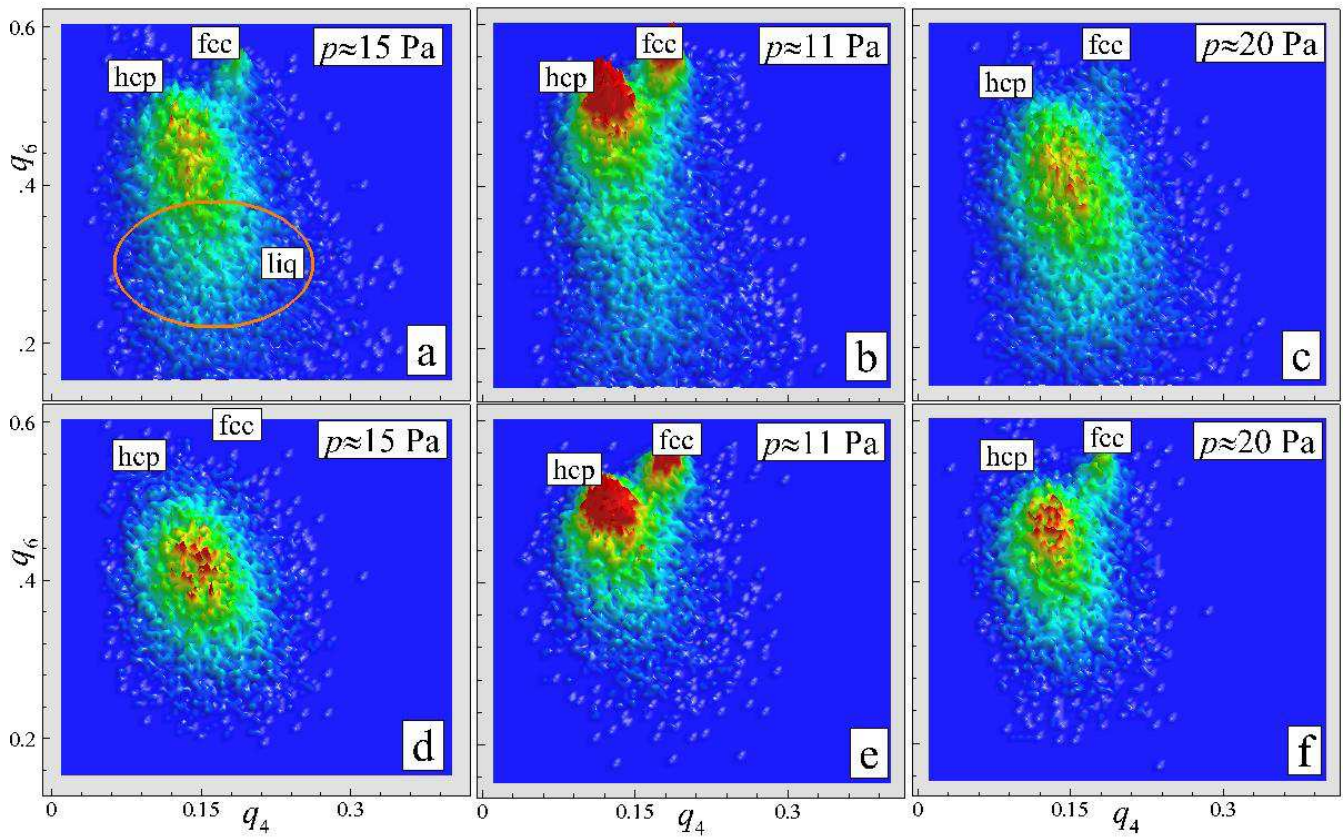}
\caption{(Color) Variation of the structural properties with pressure as reflected by particle distributions on the plane of rotational invariants ($q_4, q_6$) for the system of small (top panel) and large (bottom panel) particles. The rotational invariants for perfect hcp and fcc lattices and liquid-like domain [sketched in (a)] are also indicated. For discussion, see the text.}
\label{panel}
\end{figure*}

To determine the local structural properties of the three-dimensional particle system we use the bond order parameter method \cite{SteinPRL,SteinPRB}, which has been widely used to characterize order in simple fluids, solids and glasses~\cite{SteinPRL,SteinPRB,Wolde,ErringtonJCP}, hard-sphere (HS) systems~\cite{RintoulJCP,TorquatoPRL,RichardEPL,JinPA,KlumovAIP,KlumovPRB}, colloidal suspensions~\cite{Gasser,AuerJCP}, and, more recently, 3D complex plasmas~\cite{Milenko,Mitic,3Da,3Db,3Dc}
as well as complex plasma films~\cite{nc1,nc2}. In this method, the rotational invariants of rank $l$ of both second $q_l(i)$ and third $w_l(i)$ order are calculated for each particle $i$ in the system from the vectors (bonds) connecting its center with the centers of the $N_{\rm nn}(i)$ nearest neighboring particles:
\be
q_l(i) = \left ( \frac{4 \pi}{(2l+1)} \sum_{m=-l}^{m=l} \vert~q_{lm}(i)\vert^{2}\right )^{1/2},
\ee
\be
w_l(i) = \hspace{-0.8cm} \sum\limits_{\bmx {cc} _{m_1,m_2,m_3} \\_{ m_1+m_2+m_3=0} \emx} \hspace{-0.8cm} \left [ \bmx {ccc} l&l&l \\
m_1&m_2&m_3 \emx \right] q_{lm_1}(i) q_{lm_2}(i) q_{lm_3}(i),
\label{wig}
\ee
\noindent
where $q_{lm}(i) = N_{\rm nn}(i)^{-1} \sum_{j=1}^{N_{\rm nn}(i)} Y_{lm}({\bf r}_{ij} )$, $Y_{lm}$ are the spherical harmonics and ${\bf r}_{ij} = {\bf r}_i - {\bf r}_j$ are vectors connecting centers of particles $i$ and $j$. In Eq.(\ref{wig}) $\left [ \bmx {ccc} l&l&l \\ m_1&m_2&m_3 \emx \right ]$ denote the Wigner 3$j$-symbols, and the summation in the latter expression is performed over all the indexes $m_i =-l,...,l$ satisfying the condition $m_1+m_2+m_3=0$. The calculated rotational invariants  $q_i,~w_i$ are then compared with those for ideal lattices~\cite{SteinPRB,3Da,3Db}.
Here, we are specifically interested in identifying face-centered cubic (fcc), body-centered cubic (bcc), hexagonal close-packed (hcp), and icosahedral (ico) lattice types and, therefore, use the invariants $q_4$, $q_6$, $w_6$ calculated using the {\it fixed numbers} of $N_{\rm nn}=12$ and $N_{\rm nn}=8$ nearest neighbors, respectively.

An example from this analysis is presented in Fig.~\ref{panel}, which shows representative particle distributions on the plane ($q_4$, $q_6$). Initially, the system of small particles reveals weakly ordered fluid-like structure [Fig.~\ref{panel}(a)]. The system of large particle demonstrates more order and is apparently closer to the solid state [Fig.~\ref{panel}(d)]. Upon a decrease in pressure, the particles tend to form more ordered structures. At the minimum pressure both systems are in the solid state as can be evidenced from Figs.~\ref{panel}(b) and (e). Clear crystalline structures which are dominated by the hcp and fcc lattices are observed. Subsequent increase in the pressure suppresses the particle ordering. Figures \ref{panel} (c) and (f) demonstrate the final states of the systems, which are considerably less ordered than those in Figs.~\ref{panel}(b) and (e).

\begin{figure}
\includegraphics[width=8.2cm]{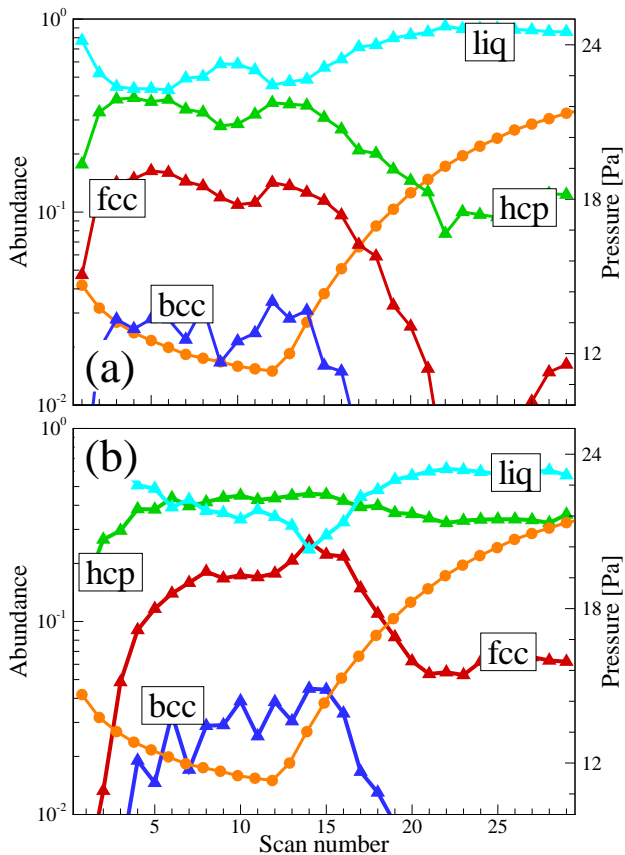}
\caption{(Color online) Structural composition of complex plasmas composed of small (a) and large (b) particles versus the scan number.
Relative number densities of hcp-like, fcc-like, bcc-like and liquid-like particles (indicated in the figure) are shown, revealing relative abundance of different phases at different pressures (pressure is shown by the orange circles on the right-hand axis).}
\label{structures}
\end{figure}

Evolution of the structural composition of complex plasmas in response to pressure manipulation is shown in Fig.~\ref{structures}. We see that the solid phase is mostly composed of hcp- and fcc-like particles with only a small portion of bcc-like clusters. A decrease (increase) in pressure enhances (suppresses) ordering of the particles. Maximum number of particles in the crystalline state clearly corresponds to the pressure range near the minimum. The system of small particles exhibits melting with increasing pressure, which is reflected by a significant drop in the number of crystalline particles. No such drop is evident for the system of large particles, indicating that it likely remains in the solid phase. In this respect we also mention that a premelting stage (disappearance of fcc- and bcc-like particles, predicted in \cite{3Db,melt_freez}) is observed in the system of small particles when the pressure increases, but is absent in the system of large particles.

\subsection{Freezing and melting indicators}

Several approximate approaches have been proposed to locate fluid-solid coexistence of various substances. This includes well known phenomenological criteria for freezing and melting, like e.g. the Lindemann melting law~\cite{Lindemann}, Hansen-Verlet freezing rule~\cite{HV}, Ravech\'{e}-Mountain-Streett criterion for freezing~\cite{RMS}, and a dynamical criterion for freezing in colloidal suspensions~\cite{DFC} (for a review see Ref.~\cite{Lowen}). These criteria are typically based on the properties of only one of the two coexisting phases and predict quasi-universal values of certain structural or dynamical quantities at the phase transition. Quasi-universality in this context means that a quantity is not exactly constant, but varies in a sufficiently narrow range for a broad variety of physical systems. It is instructive to consider application of some of these criteria to complex plasmas investigated in the present experiment.

\begin{figure}
\includegraphics [width=8.2cm] {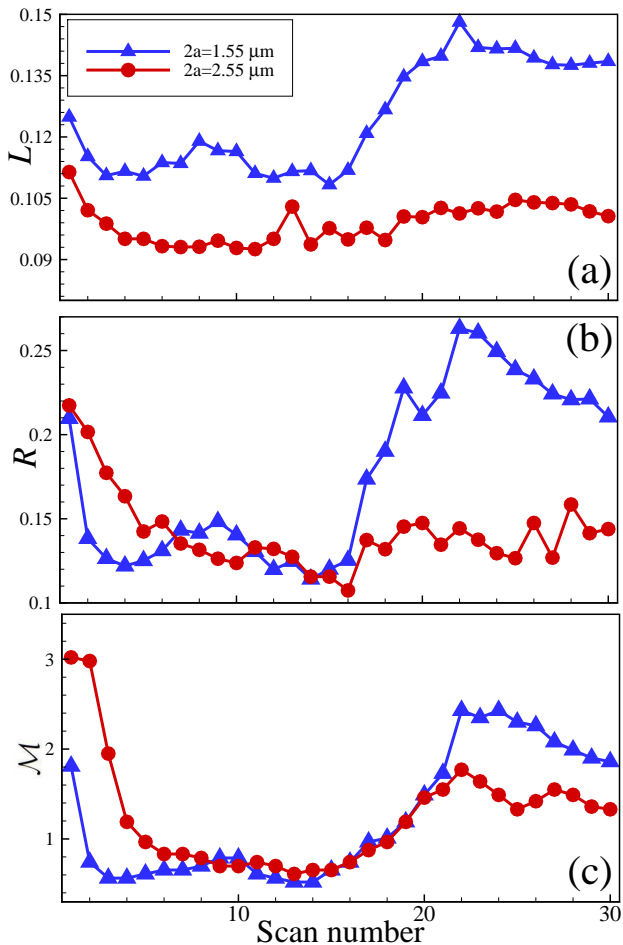}
\caption{(Color online) Comparison of various freezing and melting indicators for complex plasmas composed of small (blue triangles) and large (red circles) particles. The Lindemann measure (a), the Ravech\'{e}-Mountain-Streett ratio (b) and the melting indicator ${\cal M}$ (c) defined in the text are shown for each scan number. For discussion, see the text.}
\label{op}
\end{figure}

The Lindemann melting rule states that a crystalline solid melts when the root-mean-square displacement of particles about their equilibrium lattice positions exceeds a certain fraction of the characteristic nearest neighbor distance. The critical fraction, known as the Lindemann parameter $L$, is expected to be quasiuniversal $L\sim 0.15$. In fact, however, its exact value may depend on such factors as crystalline structure and nature (shape) of the interparticle interactions. For instance, for the inverse-power-law (IPL) family of potentials $[U(r)\propto r^{-n}]$ the values of $L$ at melting have been found to lie in the range between $\simeq 0.12$ and $\simeq 0.15$ for a wide range of $n$, $3\lesssim n\leq 100$~\cite{Agrawal}. Somewhat higher values of $L$ between $\simeq 0.15$ and $\simeq 0.18$ have been recently reported for the IPL potentials with $6\lesssim n\leq 10$, as well as for the model Gaussian and $\exp-6$ potentials~\cite{Saija}. The conventional Lindemann ratio is determined for the solid phase only. Generalizations of the Lindemann ratio which can be applied in both the solid and fluid phases have been discussed in Ref.~\cite{Chakravarty} (see also references therein). Here we define the following Lindemann-like measure convenient for the present analysis. For each particle $i$ in the system we calculate the local square deviation of the nearest neighbor distance from its (local) mean value
\begin{equation}\label{Lindemann}
\delta_i^2=N_{\rm nn}^{-1}(i)\sum_{j=1}^{N_{\rm nn}(i)}\left[r_{ij}-d_{\rm nn}(i) \right]^2,
\end{equation}
where $d_{\rm nn}(i)=N_{\rm nn}^{-1}(i)\sum_{j=1}^{N_{\rm nn}(i)}r_{ij}$ and $r_{ij}=|{\bf r}_i-{\bf r}_j|$.
The global measure is then obtained by averaging over all the $N$ particles in the system,
\begin{equation}\label{Lindemann1}
L=\sqrt{N^{-1}\sum_{i=1}^{N}[\delta_i^2/d_{\rm nn}^2(i)]}.
\end{equation}
Since the solid phase in our case is dominated by fcc and hcp lattices we find it convenient to use a fixed number of nearest neighbors, $N_{\rm nn}=12$. The behavior of the Lindemann measure defined by Eq.~(\ref{Lindemann1}) upon pressure variations is shown in Fig.~\ref{op}(a). For the system of large particles, $L$ is almost independent of the pressure and remains in the relatively narrow range $L\simeq 0.09 - 0.10$.  For the system of small particles, $L$ demonstrates similar behavior ($L\simeq 0.10 - 0.12$) for the first half of the observation sequence, but then increases considerably (up to $L\simeq 0.15$ at the maximum) in its second half (which corresponds to an increase in the neutral gas pressure). This increase in $L$ can be interpreted as a signal of melting in the system of small particles. On the other hand, almost constant values of $L$ for the system of large particles is an indication that the system remains in the solid phase. In addition, the fact that the values of $L$ are systematically smaller for the system of large particles reveals better ordering in this system. All these observations are in full qualitative agreement with the results of structural analysis described in the previous section.

The Ravech\'{e}-Mountain-Streett criterion of freezing~\cite{RMS} is based on the properties of the radial distribution function (RDF) $g(r)$ in the fluid phase. It states that near freezing, the ratio of the values of $g(r)$ corresponding to its first nonzero minimum and to the first maximum,
\begin{equation}
R=g(r_{\rm min})/g(r_{\rm max}),
\end{equation}
is constant, $R\simeq 0.2$~\cite{RMS}. This criterion describes fairly well freezing of the classical Lennard-Jones fluid, but is also not universal. For example, in studying fluid-solid coexistence of the IPL systems, the ratio $R$ at freezing was documented to vary between $\simeq 0.15$ and $\simeq 0.25$ for $n$ in the range $3\lesssim n\leq 100$~\cite{Agrawal}. Similarly, the values of $R$ in the range between $\simeq 0.18$ and $\simeq 0.24$ at freezing of the IPL, Gaussian core and $\exp-6$ model potentials have been reported~\cite{Saija}. Figure.~\ref{op}(b) shows the calculated values of the freezing indicator $R$ for different scans. Applying the threshold condition $R\simeq 0.2$ would imply that the system of small particles melts upon an increase in the neutral gas pressure (second half of the observation sequence), while the system of large particles remains in the solid state. This is consistent with the results from previous analysis.

\begin{figure}
\includegraphics[width=8.2cm]{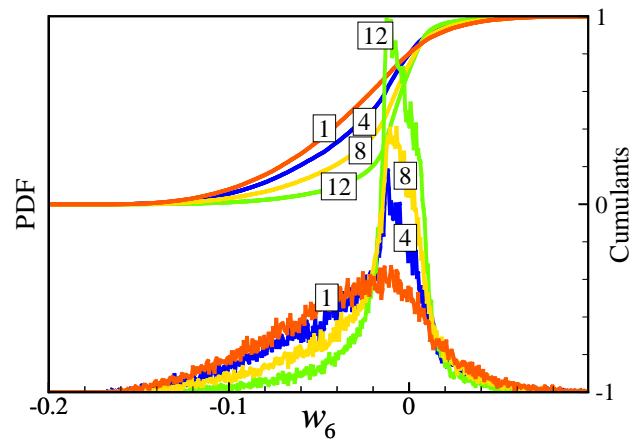}
\caption{(Color online) Distributions of small particles over the rotational invariant $w_6$ for different scan numbers (indicated in the figure). The corresponding cumulants $\hat W_6(x)$ are plotted in the upper part of the figure. The position of the half-height of $\hat W_6(x)$ can be used as a melting indicator, as discussed in the text.}
\label{w_6}
\end{figure}

It has been shown recently  that the cumulative distribution function
\begin{equation}\label{cumulant}
\hat W_6(x) \equiv \int_{-\infty}^x n(w_6)dw_6
\end{equation}
is very sensitive to the phase state of various systems \cite{melt_freez,KlumovPRB}. In equation (\ref{cumulant}), $n(w_6)$ is the distribution, normalized to unity, of particles over the rotational invariant $w_6$. Figure~\ref{w_6} shows $n(w_6)$ and the corresponding $\hat W_6(w_6)$ for four different scans corresponding to different states of the system of small particles. An appropriate  indicator of melting of the fcc solid can be defined as
\begin{equation}
{\cal M} = w_6^{\rm hh}/w_6^{\rm fcc},
\end{equation}
where $w_6^{\rm hh}$ is the position of the half-height of $\hat W_6(w_6)$ [so that $\hat W_6(w_6^{\rm hh}) =1/2$] and $w_6^{\rm fcc}=-1.3161\times 10^{-2}$ is the value of $w_6^{\rm hh}$  for the fcc lattice. Figure~\ref{op}(c) shows the values of ${\cal M}$ calculated for each scan number. In Ref.~\cite{CrystPRL} we used the threshold value ${\cal M} \simeq 1.3$ at melting. For this value, however, both the systems of small and large particles appear to melt upon increase in the pressure [see Fig.~\ref{op}(c)]. If we require that only the system of small particles exhibits melting (in agreement with the other two phase change indicators discussed above), then the ``real'' melting threshold value should be somewhat increased (to ${\cal M} \simeq 1.5 - 2.0$). Qualitatively, the ${\cal M}$-based measure adequately describes the decay of ordering accompanying an increase in the neutral gas pressure.

\section{Theoretical Interpretation}

\subsection{Plasma parameters and particle charges}

We attribute the observed fluid-solid phase change in the system of small particles to the variation in the electrical repulsion between the particles. Manipulating the gas pressure experimentally, changes various complex plasma parameters and modifies the strength of the repulsion. When the electrical coupling reach (or drops below) certain level, freezing (or melting) occurs. To verify this scenario we need to estimate relevant complex plasma parameters.

We use the results from SIGLO-2D simulations \cite{PK3+} to estimate plasma parameters in the absence of particles. Earlier, it has been shown that SIGLO simulation yields reasonable agreement with the results from Langmuir probe measurements for a similar discharge chamber~\cite{Klindworth}. In the considered regime ($p \sim 10-25$ Pa, rf-amplitude $\sim 15$ V) the central plasma density is linear on $p$ and can be, with a reasonable accuracy, described as $n_0\simeq (1.20+0.11p)\times 10^8$, where $n_0$ is in cm$^{-3}$ and $p$ in Pa. The electron temperature exhibits almost no dependence on pressure, $T_e\simeq 3.8$ eV. Ions and neutrals are at room temperature $T_{i,n}\sim 0.03$ eV.

When particles are injected into the discharge they inevitably modify plasma parameters. In the following we assume that {\it inside the particle cloud} the electron temperature remains unaffected, while the electron and ion densities are modified to keep quasineutrality, $n_e+|Q/e|n_p\simeq n_i$, where $Q$ is the particle charge. Furthermore, we assume that $n_e$ remains close to the particle-free value $n_0$, while $n_i$ somewhat increases in response to perturbations from the particle component. Physically, this assumption corresponds to the case of static equilibrium, when electron and ion densities satisfy Boltzmann relation in the local plasma potential modified by the presence of charged dust particles. It is in reasonable agreement with numerical simulation results~\cite{Land,Goedheer} regarding plasma parameters {\it inside the particle cloud} (simulations show, however, that {\it inside the void region} in the center of the discharge, the plasma density and electron temperature can be considerably higher than those in the particle-free discharge \cite{Land,Goedheer}). Note that in the opposite limiting case, when the ion equilibration length is much longer than the characteristic size of the particle cloud (e.g. sufficiently thin cloud), it would be more reasonable to assume that the ion density is constant, whilst electron density is depleted~\cite{Havnes}. Such a situation has been apparently realized in a recent experiment~\cite{Goertz} at a very low neutral gas pressure.

\begin{figure}
\includegraphics[width=8.2cm]{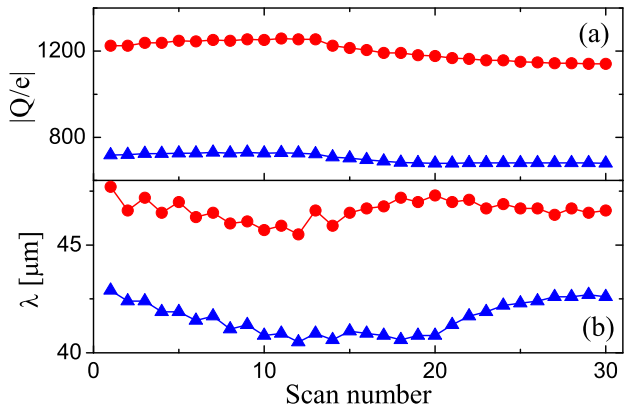}
\caption{(Color online) Estimated values of the particle charge number $|Q/e|$ (a) and effective plasma screening length $\lambda$ (b) vs. the scan number. Red circles (blue triangles) correspond to the system of large (small) particles.}
\label{parameters}
\end{figure}

Using these assumption we can calculate the dependence of the particle charge on pressure (and interparticle distance) employing certain model for the electron and ion fluxes absorbed on the particle surface and requiring that these fluxes balance each other. For the electron flux we use the orbital motion limited (OML) theory~\cite{Mott,Allen}
\begin{equation}\label{Je}
J_e=\sqrt{8\pi}a^2n_ev_{T_e}\exp(-z),
\end{equation}
where $z=|Q|e/aT_e$ is the reduced particle charge and $v_{T_e}=\sqrt{T_e/m_e}$ is the electron thermal velocity. For the ions, however, the OML approximation would be much less appropriate. There are two main reasons for that. First, as has been evidenced in a number of recent studies~\cite{Book,Zobnin,Lampe,Hutchinson,Zobnin1,FortovExperiment,RatynskaiaPRL,KhrapakPRE,KhrapakCPP,KhrapakEPL}, ion-neutral collisions can represent a very important factor affecting and regulating the ion flux collected by the particle. A collision between an ion and a neutral atom lowers the ion energy (in this respect resonant charge exchange collisions are especially important) and destroys its angular momentum. If such a collision occurs sufficiently close to the negatively charged particle, an ion which experienced a collisions has a very low probability to overcome the attraction from the particle and will eventually reach its surface. Thus, in the considered weakly collisional regime, ion-neutral collisions result in a more effective ion collection by the particles, i.e. they increase the ion flux. The second factor, discussed very recently in the context of particle charging in plasmas, is associated with the ionization processes. Similarly to ion-neutral collisions, electron impact ionization events create slow ions  in the vicinity of the particle and therefore enhance the ion flux towards its surface~\cite{Ionization}.

To make a quantitative estimate we employ an expression of Lampe {\it et al}.~\cite{Lampe} for the collision-enhanced ion flux
\begin{equation}\label{Lampe}
J_{\rm ic} \simeq \sqrt{8\pi}a^2 n_i v_{T_i} \left[1+z\tau+ (R_0^3/a^2\ell_i)\right],
\end{equation}
where $v_{T_i}=\sqrt{T_i/m_i}$ is the ion thermal velocity, $\tau=T_e/T_i$ is the electron-to-ion temperature ratio, $\ell_i$ is the ion mean free path with respect to collisions with neutrals ($\ell_i\simeq T_n/p\sigma_{in} $, where $\sigma_{in}\simeq 2\times 10^{-14}$ cm$^2$ is the effective ion-neutral collisions cross section in argon), and $R_0$ determines the radius of a sphere around the particle, inside which the potential energy of ion-particle interaction is sufficiently high (higher than the average kinetic energy of an ion after a collision). The first two terms in the square brackets on the right-hand side of Eq.~(\ref{Lampe}) reproduce the result of the collisionless OML model. The third term represents the collision-driven flux enhancement. There is no consensus presently regarding the precise determination of $R_0$ in this collisional term and different arguments have been put forward~\cite{Lampe,KhrapakPRE,KhrapakEPL,Tolias}. In general, $R_0$ should depend on the exact shape of the ion-particle interaction potential, details of the ion-neutral collisional processes, distribution of ion velocities, etc. In a recent paper \cite{KhrapakEPL} it has been suggested that in the regime of sufficiently small particles and weak ion-particle coupling, the Coulomb radius $R_{\rm C}=|Q|e/T_i=az\tau$ can be used as a relevant measure of $R_0$. This choice has been shown to agree reasonably well with the experimental results for very small ($1.31$ $\mu$m in diameter) particles~\cite{KhrapakEPL}. In the present experiment, however, the particles sizes are somewhat larger, and the applicability condition to set $R_{0}$ equal to $R_{\rm C}$ (which requires $R_{\rm C}\lesssim \lambda$), is violated, at least for larger particles. To get an idea what kind of modification is required, let us consider the opposite limit of strong ion-particle coupling, $R_{\rm C}\gg \lambda$. In this case, assuming the Debye-H\"{u}ckel (Yukawa) form of the interaction potential, we can easily estimate the length scale of the region of strong ion-particle interaction as $R_0\simeq \lambda\ln(R_{\rm C}/\lambda)$~\cite{KhrapakPRL2003}. A simple expression of the type $R_0\simeq \lambda\ln(1+R_{\rm C}/\lambda)$ would therefore describe adequately the corresponding limits of weak and strong ion-particle coupling and provide a smooth transition between them. We adopt this heuristic approximation in the analysis below.

Regarding the effect of ionization, in Ref.~\cite{Ionization} it has been demonstrated that the relative magnitude of ionization and collisional enhancements of the ion flux is roughly given by the ratio of the corresponding frequencies, $\nu_{\rm I}/\nu_{\rm in}$, where $\nu_{\rm I}$ is the ionization frequency and $\nu_{\rm in}\simeq v_{T_i}/\ell_i$ is the frequency of ion-neutral collisions.
This implies that the effects can be added in a simple superposition, which yields the following expression for the ion flux
\begin{widetext}
\begin{equation}\label{Ji}
J_i\simeq \sqrt{8\pi}a^2 n_i v_{T_i} \left[1+z\tau+ \left(1+\nu_{\rm I}/\nu_{\rm in}\right)\left(\lambda^3/a^2\ell_i\right)\ln^3\left(1+R_{\rm C}/\lambda\right)\right] .
\end{equation}
\end{widetext}
For a fixed gas and ion temperature, the ratio $\nu_{\rm I}/\nu_{\rm in}$ is a function of a single parameter -- electron temperature $T_e$. The corresponding function has been evaluated for neon and argon plasmas with room temperature ions~\cite{Ionization}. From the results shown in Fig. 1 of Ref.~\cite{Ionization} we conclude that  $\nu_{\rm I}/\nu_{\rm in}\simeq 1.8$ at $T_e\simeq 3.8$ eV.

The charge can then be estimated from the balance condition $J_i=J_e$ using expressions (\ref{Je}) and (\ref{Ji}). In doing so we take into account the modifications of the ion density compared to the particle-free value discussed above. Quasineutrality condition implies  $n_i/n_e=n_i/n_0\simeq 1+zP$, where $P=(aT_e/e^2)(n_p/n_0)$ is the scaled particle-to-plasma density ratio (the so-called Havnes parameter). The effective screening length (screening is mostly associated with the ion component for $\tau\gg 1$) is $\lambda=\lambda_0/\sqrt{1+zP}$, where $\lambda_0=\sqrt{T_i/4\pi e^2n_0}$ is the unperturbed ion Debye radius.

The resulting dependence of the particle charge and plasma screening length on the scan number is shown in Fig.~\ref{parameters}. We observe that the absolute magnitude of the particle charge (circles) slightly decreases with increasing pressure while the plasma screening length (triangles) exhibits the opposite behavior. The first tendency is expected: Increase in the collisionality lowers the absolute magnitude of the particle charge in the weakly collisional regime~\cite{Book,KhrapakCPP}. However, the effect is much less pronounced than it would be for an individual particle. This is the result of the coupling between $n_e$, $n_i$, and $|Q/e|n_p$. Namely, a decrease in $n_p$ with increase in the pressure results in a decrease in $n_i$ and, hence, in $J_i$, so that the collisional enhancement of the ion flux is almost compensated by ion depletion in the considered regime. The dependence of the screening length on pressure is unexpected. In the particle-free plasma one would expect approximately $\lambda\propto n_i^{-1/2}\propto p^{-1/2}$, i.e. a decrease of $\lambda$ with increasing pressure. Here again, a decrease in the ion density triggered by the expansion of the particle cloud with increasing $p$ provides overcompensation and the trend reverses. Overall, the relative variations in $Q$ and $\lambda$ are rather weak, considerably weaker than those in the interparticle distance.

Thus, the coupling between $n_e$, $n_i$, and $|Q/e|n_p$ via the quasineutrality condition and the the corresponding effect of charge reduction in dense particle clouds~\cite{Havnes,Goertz,Barkan} plays essential role in the present experiment. Note that this effect is to some extent similar to a reduction of colloidal charge when increasing the colloidal volume fraction. It has been demonstrated that as a result of this charge reduction colloids can exhibit an intriguing reentrant melting behavior, when a colloidal fluid phase appears at a higher volume fraction than a colloidal crystal~\cite{Royall,Smallenburg}. Similar behavior -- reentrant fluid-solid-fluid series of phase changes upon isothermal compression -- can be expected in complex plasmas, as discussed recently in Ref.~\cite{KhrapakEPL2010}. However, the particle clouds observed in the present experiment are not dense enough to exhibit such a behavior. As we have seen, in the regime investigated, some increase in $\Delta$ with pressure is accompanied by almost constant values of the particle charge and screening length.

\subsection{Interaction and coupling strength}

The pairwise potential of electrical interaction between highly charged particles is often assumed to be of Debye-H\"{u}ckel (Yukawa) form,
\begin{equation}\label{Yukawa}
U(r)=(Q^2/r)\exp(-r/\lambda),
\end{equation}
where $r$ is the interparticle distance. The actual interactions are known to be considerably more complicated especially at large interparticle separations. In particular, continuous plasma absorption on the particle surface can give rise to unscreened inverse-power-law long-range asymptotes of the potential~\cite{KhrapakCPP,TsytovichUFN,Allen2000,Filippov2007,KKM,ChaudhuriIEEE}. In addition, plasma openness, associated with constant plasma absorption on the particles, can give rise to the so-called ``ion-shadowing'' attraction~\cite{TsytovichUFN,Ignatov,LampePoP2000,KhrapakPRE2001,CP}. It basically represents the plasma drag that one particle experiences as a consequence of the plasma flux directed to another neighboring particle and vice versa. Electron and ion production (ionization) and loss (e.g. recombination) can result in the emergence of the two dominating exponentially screened asymptotes (a double-Yukawa repulsive potential), one of which is determined by the classical mechanism of Debye-H\"{u}ckel screening, while the other is merely controlled by the balance between the plasma production and loss~\cite{FilippovJETP,KhrapakPoP2010}. Here we neglect all these corrections and adopt the simple form (\ref{Yukawa}). The justifying arguments are as follows: (i) The mean interparticle separation is not large enough in the present experiment ($\Delta\sim 3\lambda$) for the
IPL asymptotes to dominate~\cite{KKM}; (ii) The neutral gas pressures used in this experiment are well above those required to make ``ion shadowing'' attraction operational~\cite{KM2008}; (iii) The relative ionization efficiency is not high enough to expect significant deviations from the conventional screening regime (\ref{Yukawa}) at distances characterizing the mean interparticle separation~\cite{KhrapakPoP2010}.

\begin{figure}
\includegraphics[width=8.2cm]{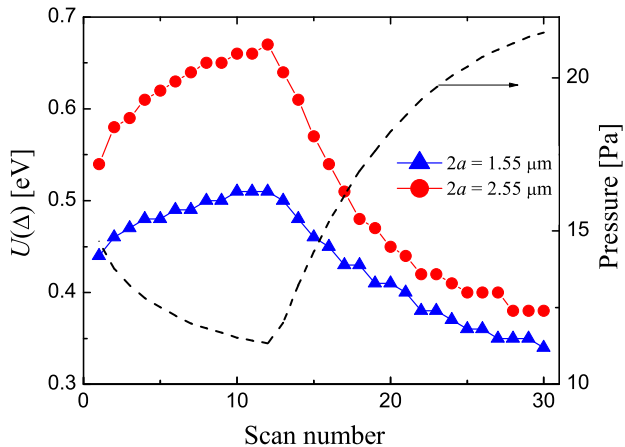}
\caption{(Color online) Estimated interparticle interaction energy $U(\Delta)$ at each scan. Blue triangles correspond to a complex plasma composed of small particles. Red circles are for the system of large particles. The dashed curve shows the evolution of the neutral gas pressure in the experiment. Note that the interaction energy is more than one order of magnitude higher than the gas (room) temperature ($T_n\simeq 0.03$ eV), indicating that both systems always remains in the strongly coupled state.}
\label{energy}
\end{figure}

Using the measured values of $\Delta$ and estimated values of $Q$ and $\lambda$ we can evaluate variations in the interaction energy between neighboring particles $U(\Delta)$.
The results are shown in Fig. \ref{energy}. The energy, as a function of $p$, exhibits maximum at the lowest pressure $p\simeq 11$ Pa in both cases studied. Thus, increasing (decreasing) the neutral gas pressure implies reduction (grows) in the strength of electrical repulsion (coupling strength) between the particles. As discussed above, the main factor responsible for this behavior is the compression (expansion) of the particle system upon decrease (increase) in the pressure ($Q$ and $\lambda$ are less sensitive to pressure manipulations). Consequently, complex plasma appears more ordered when the pressure is low and less ordered when the pressure is high (in the pressure range investigated). Freezing and melting can also be expected, provided the system is located not too far from the phase boundary. This is in full qualitative agreement  with the results from the structural analysis.

\subsection{Equilibrium Phase Diagram}

To get further insight into phase behavior of complex plasmas under investigation, let us estimate the location of phase trajectories of these system on a relevant {\it equilibrium} phase diagram.

The system of particles interacting via the Yukawa potential (\ref{Yukawa}) in thermodynamical equilibrium can be characterized by two dimensionless parameters. In the field of complex (dusty) plasmas the {\it screening parameter} $\kappa=\Delta/\lambda$, characterizing the efficiency of screening, and the {\it coupling parameter} $\Gamma=(Q^2/T_p\Delta)\exp(-\kappa)$, measuring the ratio of the interaction energy at the mean interparticle separation to the particle kinetic temperature, are commonly used. Equilibrium phase diagrams of Yukawa systems have been extensively studied~\cite{Kremer1986,Robbins1988,Meijer1991,Stevens1993,Hamaguchi1997}. In the strongly coupled regime they can exist in the fluid or solid phases. In the solid phase particle form either bcc (weak screening regime) or fcc (strong screening regime) lattices. The triple point is located at $\kappa\simeq 6.9$ and $\Gamma\simeq 3.5$~\cite{Hamaguchi1997} (according to a more recent estimate from Ref.~\cite{Hoy} its location is at $\kappa\simeq 7.7$ and $\Gamma\simeq 3.1$). The boundary between the fluid and solid phases on the plane ($\kappa$, $\Gamma$) can be approximated by the expression~\cite{VaulinaJETP,VaulinaPRE}
\begin{equation}\label{melting}
\Gamma_{\rm M}\simeq \frac{106}{1+\kappa+\tfrac{1}{2}\kappa^2},
\end{equation}
where the subscript ``M'' refers to melting (the density gap is rather small so that it makes little sense to distinguish between freezing and melting here). Other approximate analytical expressions have been also proposed in the literature to locate the fluid-solid coexistence for a wide class of interaction potentials~\cite{KM_2009,Principle,PrincipleAIP}. We use Eq.~(\ref{melting}) here due to its particular simplicity and reasonable accuracy: Deviations between the results from Eq.~(\ref{melting}) and numerical simulation data from Ref.~\cite{Hamaguchi1997} do not exceed several percent, as long as $\kappa$ remains not too large ($\kappa\lesssim 8$)~\cite{VaulinaJETP}.

\begin{figure}
\includegraphics[width=7.8cm]{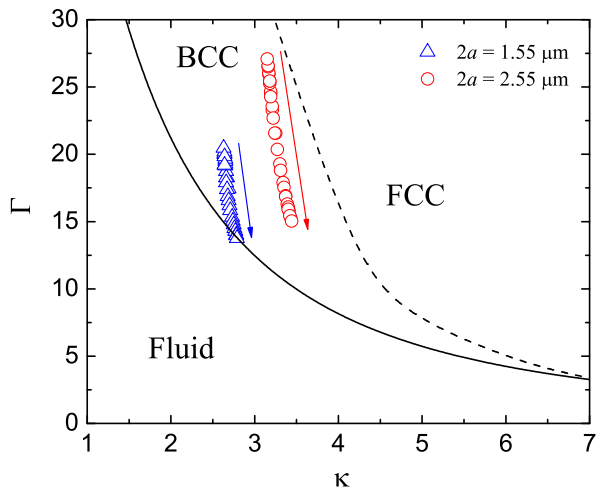}
\caption{(Color online) Estimated trajectories of complex plasmas, studied in the present experiment, on the equilibrium phase diagram of Yukawa systems. The solid curve corresponds to the fluid-solid phase change [Eq.~(\ref{melting})], the dashed curve shows the boundary between the bcc and fcc solids (smooth fit to the simulation data points from Ref.~\cite{Hamaguchi1997}). Blue triangles denote phase state points for the system of small particles, red circles correspond to large particles. Arrows mark the direction of the phase evolution when the pressure increases.}
\label{PD}
\end{figure}

Figure \ref{PD} shows the equilibrium phase diagram of Yukawa systems in the ($\kappa$, $\Gamma$) plane along with the estimated phase-state points from the present experiment. In this figure circles and triangles correspond to the phase states visited by the system of small and large particles, respectively. The arrows indicate the direction of phase evolution when the pressure increases. With increasing pressure both systems move towards the melting curve. Complex plasma composed of small particles crosses the phase boundary for highest pressures investigated. At the same time, complex plasma formed by large particles always remains in the solid state. These result are  in reasonable quantitative agreement with those from structural order measurements and estimated values of freezing/melting indicators. On the other hand, there is a notable disagreement with respect to the structure of the solid phase. Equilibrium Yukawa crystals, in the range of $\kappa$ investigated, should form the bcc lattice (see Fig.~\ref{PD}). In contrast, the observed crystalline structures are dominated by the hcp and fcc lattices (see Fig.~\ref{structures}). Such a drastic discrepancy can be associated with the fact the crystallization evolves over several essentially non-equilibrium stages, yet the physics of these non-equilibrium processes is still poorly understood. In particular, this concerns the crystal nucleation -- both homogeneous and heterogeneous -- out of the supercooled melt (when the thermodynamically stable solid has a structure which
is vastly different from that of the liquid)~\cite{Auer2004}, as well as the crystal nucleation out of a thermodynamically unstable solid phase (such as hcp and fcc lattices, which have a structure incompatible with the ground-state bcc crystal). In this case, one could expect largely prevented nucleation of the equilibrium phase, and our experiment might be a manifestation of
this effect.

\section{Discussion}

In the first publication related to the discussed experiment~\cite{CrystPRL} the theoretical interpretation was somewhat different. In particular, we did not take into account the effect of ionization enhanced ion collection by the particle. In this way, the particle charge was apparently overestimated. As a result, we had to assume that the particle kinetic temperature $T_p$ was several times higher than the neutral gas temperature in order to explain melting. Estimates presented in the present paper yield lower charges and, therefore, make such an assumption unnecessary (see Fig.~\ref{PD}).  Unfortunately, the particle dynamics is not resolved in this experiment, so we cannot completely exclude the possibility that $T_p$ is somewhat higher than $T_n$.

It is instructive to discuss some specific properties of the phase transitions observed. As pointed out above, an increase in $\Delta$ (i.e. decrease in the particle density) is the main factor responsible for the melting when the neutral gas pressure increases. Similarly, reducing pressure compresses the particle system and this stimulates freezing. This is a generic mechanism of (isothermal) fluid-solid phase transition, that can be realized in extremely wide range of various substances and materials. In complex plasmas, it can be in principle observed in both 3D and 2D systems~\cite{Knapek1,Nosenko}. However, 2D complex plasmas are known to be strongly affected by plasma-specific mechanisms of melting. One of the clear manifestations is the conventional procedure of melting flat plasma crystals by {\it reducing} the neutral gas pressure in ground-based experiments~\cite{Morfill_Nat,Melzer}. The difference is not a consequence of the essentially 2D character of crystals investigated on Earth, but is rather due to the presence of strong electric fields (and, therefore, strong ion flows) required to balance the force of gravity. There are effective mechanisms of converting the energy associated with ion flows into the kinetic energy of the particles. Known scenarios include ion-particle two-stream instability~\cite{Glenn,Ganguli}, non-reciprocity of the interaction due to asymmetric character of the screening cloud around the particles (``plasma wakes'')~\cite{Couedel,Schweigert,IvlevPRE2001}, particle charge variations~\cite{Nunomura,Ivlev}. All these scenarios lead to an abrupt increase of the particle kinetic energy at pressures below certain threshold value, causing crystal melting. The process of melting observed in the present experiment is apparently free of these plasma-related effects, but has much more in common with generic processes in conventional atomic, molecular, and soft matter systems.

\section{Conclusions}

We have presented a comprehensive overview of the experimental studies on fluid-solid phase transitions in large 3D complex plasma clouds performed in microgravity conditions onboard the ISS. The neutral gas pressure turns out to be a convenient control parameter to drive crystallization and melting. In the parameter regime investigated, the phase transition is mostly associated with the compression (expansion) of the complex plasma system upon decrease (increase) in pressure, associated with the variations in plasma confinement. This is very different from the conventional procedure of melting flat (2D) complex plasma crystals by reducing the gas pressure in ground-based experiments. Detailed analysis of complex plasmas structural properties and evaluation of three different freezing/melting indicators reveal an overall good qualitative agreement. Theoretical estimates of the complex plasma parameters allow us to approximately determine the locations of the systems investigated on the relevant (Yukawa systems) equilibrium phase diagram. Here again, reasonable agreement is documented, except the structure of the solid phase, which is expected to be bcc lattice form numerical simulations, but is dominated by the hcp and fcc domains in experiment. We attribute this to the non-equilibrium character of phase evolutions studied in the present experiment, but this issue certainly deserves more attention. Further investigations are planned, which we believe will provide important insight regarding the principle mechanisms dominating ubiquitous and still poorly understood phenomena of crystallization and melting in complex plasmas and related systems of strongly coupled particles.

\section{Acknowledgments}

This work was supported by DLR under Grants 50WP0203 and 50WM1203.
(Gef\"{o}rdert von der Raumfahrt-Agentur des Deutschen Zentrums f\"{u}r Luft und Raumfahrt e. V. mit Mitteln des Bundesministeriums f\"{u}r Wirtschaft und Technologie aufgrund eines Beschlusses des Deutschen Bundestages unter dem F\"{o}rderkennzeichen 50 WP 0203 und 50 WM 1203). We additionally acknowledge financial support from the European Research Council under the European Union's Seventh
Framework Programme (FP7/2007-2013) / ERC Grant agreement 267499.

\end{document}